\documentclass{article}
\usepackage{spconf,amsmath,graphicx}
\usepackage{caption}
\usepackage{subcaption}
\usepackage{tabularx}
\usepackage{array}
\usepackage{algorithm}
\usepackage{algpseudocode}
\newtheorem{definition}{Definition}

\newtheorem{remark}{Remark}

\title{MULTIVARIATE SIGNAL DENOISING BASED ON GENERIC MULTIVARIATE DETRENDED FLUCTUATION ANALYSIS}
\vspace{-5mm}
\name{Khuram Naveed $^a$, Sidra Mukhtar $^a$ and Naveed ur Rehman $^b$}
\address{$^a$\normalsize{Department of Electrical and Computer Engineering, COMSATS University Islamabad (CUI), Islamabad, Pakistan}\\ 
	$^b$\normalsize{Department of Electrical and Computer Engineering, Aarhus University, Aarhus, Denmark}}

\begin{document}
%\ninept
\topmargin=0mm
\maketitle
%\vspace{-5mm}
\begin{abstract}
We propose a novel multivariate signal denoising method that performs long-range correlation analysis of multiple modes in input data by considering inherent inter-channel dependencies of the data. That is achieved through a novel and generic multivariate extension of detrended fluctuation analysis (DFA) method - another contribution of this paper. Specifically, our proposed denoising method first obtains data driven multiscale signal representation using multivariate variational mode decomposition (MVMD) method. Then, the proposed generic multivariate DFA is used to reject the predominantly noisy modes based on their randomness scores. Finally, the denoised signal is reconstructed by summing the remaining modes albeit after the removal of the noise traces using the principal component analysis (PCA).
\end{abstract}
\vspace{-3mm}
\section{Introduction}
\label{sec:intro}
\vspace{-3mm}
Multi-sensor systems have found widespread use in many applications including medical diagnosis, health monitoring, weather forecasting etc. Within these systems, a network of synchronized sensors is used to record signals originating from physical system(s) resulting in interdependent multichannel observations. Those observations, denoted by $\pmb{x}_i\in \mathcal{R}^m$, are modelled as a combination of the desired signal $\pmb{s}_i \in \mathcal{R}^m$ and the unwanted noise $\boldsymbol{\psi}_i=\mathcal{R}^m$, as follows
%which, in its raw form, are composed of noise along with the desired signal, as follows
% contain both desired signal $\pmb{s}_i\in \mathcal{R}^m$ and the undesired noise parts, as follows
\vspace{-2mm}
\begin{equation}\label{Eq01}
	\pmb{x}_i=\pmb{s}_i + \boldsymbol{\psi}_i, \ \ \forall \ \ i=1,\ldots,N.
\end{equation}
%where $\pmb{s}_i\in \mathcal{R}^m$ denotes a desired multivariate signal observation at time index $i$, while $\boldsymbol{\psi}_i\in \mathcal{R}^m$ denotes a multivariate additive noise observation. 
%which is assumed to be independent and identically distributed by zero mean multivariate Gaussian distribution $\mathcal{N}_m(\mathbf{0},\Sigma)$ 
%with covariance matrix $\Sigma$.  
\vspace{-7mm}

%Several methods have been proposed in literature to 
Estimation of true multivariate signal $\pmb{s}_i$ from raw signal recordings $\pmb{x}_i$ is a problem of considerable interest. To solve this problem, most of the existing algorithms are direct multichannel extensions of the popular multiscale approaches that have worked extremely well on univariate (single-channel) data.
%Consequently, the acquired raw data is pre-processed using a denoising method that minimizes the undesired noise to an extent that important signal features remain intact. Traditionally, multiscale representation of multivariate data is put to good use in this regard. 
For instance, the sparsity of discrete wavelet transform (DWT) is exploited to reject noise via a multichannel expansion of the universal threshold \cite{donoho1995Visu}. 
%Just recently, in \cite{naveed2020MGWD}, a novel multivariate GoF test is introduced for multichannel denoising. 
%Similarly, \cite{ahrabian2015MWSD} utilizes multichannel synchrosqueezing wavelet transform for this purpose. Apart from that, multivariate empirical mode decomposition (MEMD) \cite{rehman2009MEMD,rehman2015dynamicMEMD} has also been used for multivariate signal denoising using interval thresholding \cite{hao2017MEMD-IT,ur2019MMD}. 
Similarly, multiscale denoising approaches for multivariate data that are based on synchrosqueezed wavelet transform \cite{ahrabian2015MWSD}, multivariate empirical mode decomposition (MEMD) \cite{hao2017MEMD-IT,ur2019MMD} and translation invariant DWT aided by Mahalanobis distance measure \cite{naveed2020MGWD}, are extensions of \cite{meignen2012UWSD,kopsinis2009EMD-IT,ur2017DWTGoF,ur2016GoFImageConf} respectively. 
%have also been popular.
%where  employs a joint of univariate interval thresholding function (originally introduced in \cite{kopsinis2009EMD-IT}) while \cite{ur2019MMD} performs multivariate interval thresholding using Mahalanobis distance that operates simultaneously on all channels. 
Moreover, variational mode decomposition (VMD) algorithm \cite{dragomiretskiy2014VMD} and its multivariate extension \cite{ur2019MVMD} have been employed for denoising \cite{liu2016VMDDFA,naveed2021VMD-CVM,cao2020MVMD-HDSP}.
%have emerged as more powerful new tools for nonstationary signal analysis. 
%as compared to EMD/ MEMD, owing to their sound mathematical foundation. 
%Consequently, 
%MVMD has been used for multivariate signal denoising in \cite{cao2020MVMD-HDSP}. 
%Hausdoff distance has been used in combination with MVMD for multichannel signal denoising. 
%In the context of univariate data, a more rigorous approach employs 
%these class of methods have been used for data denoising both for single-channel \cite{liu2016VMDDFA} and multichannel data sets \cite{cao2020MVMD-HDSP}. 
In \cite{liu2016VMDDFA}, detrended fluctuation analysis (DFA) \cite{peng1994DFA} has been used to identify and reject the signal modes with predominant noise by estimating 
their long-range correlations.

\vspace{-0.15mm}
In its original form, DFA only caters for single-channel time series data. While its multichannel extension exists \cite{xiong20171stMDFA}, it processes each data channel in isolation thereby ignoring inter-channel 
correlations within multivariate data.  
To that end, we first develop a novel and generic multichannel extension of DFA, termed GMDFA in the sequel, that fully incorporates inter-channel correlations within data using Mahalanobis distance. Then, using that extension, we present a novel multichannel multiscale denoising method that first uses MVMD to decompose a multivariate data into multiple frequency modes; and then identifies (and rejects) the noisy modes using GMDFA. The efficacy of the proposed approach is demonstrated on a variety of real multichannel signals.  
\vspace{-2mm}
\section{Detrended Fluctuation Analysis}
\vspace{-2mm}
%Generally, Hurst exponent is used to estimate the extent of long-range correlations in a time series but
%%whether the time series is short-range correlated (i.e., $0<H<0.5$), completely uncorrelated (i.e., $H=0.5$) or long-range correlated (i.e., $0.5<H<1$) \cite{}. 
%its widespread use is restricted by the spurious auto-correlation scores in case of nonstationary data \cite{}. 
The detrended fluctuation analysis (DFA) is widely used to estimate the extent of long-range correlations in a nonstationary time series. The main advantage of using DFA is that it circumvents the artefacts of nonstationarity (e.g., local trend, noise etc.,) which cause spurious scores in the otherwise used Hurst exponent method \cite{peng1994DFA}.
%circumvent the spurious scores by removing the artifacts via the detrending operation.  
Specifically, DFA estimates a power law \textit{scaling exponent} 
%while also indicating about other statistical properties, e.g., degree of smoothness of a time series 
by observing natural variability of signal fluctuations around its local trend at different time scales. As a result, intrinsic fluctuations of a time series are extracted by \textit{detrending} the slowly oscillating background that causes spurious scores \cite{peng1994DFA} as described below:

Given a time series $x_i, \ \forall \ i=1,...,N$; its normalized cumulative sum is obtained as follows: $\small{y_i = \frac{1}{N}\sum_{i=1}^N (x_i-\overline{x})}$, where $\overline{x}$ denotes the signal mean.
%after each value $x_i$ is shifted by their mean $\overline{x}$.$\small{y_i = \frac{1}{N}\sum_{i=1}^N (x_i-\overline{x})}$
%$y_i = \frac{1}{N}\sum_{i=1}^N (x_i-\overline{x})$
%\begin{equation}
%\small{y_i = \frac{1}{N}\sum_{i=1}^N (x_i-\overline{x}),}
%\end{equation}
The resulting profile $y_i$ is then divided into $N_s = N/s$ segments of equal length $s$ from both ends. Next, least squares polynomial fitting approach is employed on the resulting segments to estimate the local trend, denoted by $\tilde{y}_i$. Finally, a root mean squared (RMS) function $F(s)$ of detrended fluctuations $y_i-\tilde{y}_i$ is obtained as
%by taking the RMS of the detrend, i.e., error between the profile $y_i$ and the fit $\tilde{y}_i$,
\begin{equation}\label{Eq02}
	\small{F(s) = \sqrt{\frac{1}{2N_s}\sum_{v=1}^{2N_s} \left(\frac{1}{s}\sum_{i=1}^{s}\left(y_i-\tilde{y}_i\right)^2\right)}.}
\end{equation}

Note from \eqref{Eq02} that $F(s)$ is the root mean of local (segment) variances that is expected to increase with increase in the time scale $s$. This increase in $F(s)$, when described using the power law relation of the time scale $s$ reflects on the long range correlations of a time series \cite{kantelhardt2001detecting}. 
% based on the power law relation between the time scale $s$ the.  Hence, the analysis of how power law relation of $F(s)$ with change in time scale $s$ reveals  
%\begin{equation}\label{Eq02}
%\small{F(s) = s^\alpha},
%\end{equation}
%where $\alpha$ denotes the long-range correlation exponent  
Specifically, the scale exponent $\alpha$ indicates long-range correlations if $\alpha>0.5$; while the cases of $\alpha=0.5$ and $\alpha<0.5$ suggest no-correlations and short-range correlations, respectively. 
Furthermore, $\alpha$ informs about the degree of smoothness of a time series, i.e., a higher value $\alpha$ indicates the presence of slow fluctuations while a lower $\alpha$ hints at rapid fluctuations \cite{mert2014EMD-DFA1}. 
%Therefore, DFA is relevant in a variety of applications involving
The resulting insight gained through DFA renders it suitable in many signal processing related applications involving signal analysis \cite{leistedt2007DFA_Application} and denoising \cite{liu2016VMDDFA}.
%	mert2014EMD-DFA2}.

%Given a multivariate time series $\pmb{x}_i \ \forall \ i=1,...,N$ such that $\pmb{x}_i=[x_{i_1},\ldots,x_{i_m}]\in\mathcal{R}^m$ denote $m$-variate observations. Let $y_{i_n}$ be the cumulative sum corresponding to each channel $x_{i_n}$ indexed by $n$. 
A multichannel DFA is presented in \cite{xiong20171stMDFA} using a straightforward multichannel generalization of \eqref{Eq02} which is given by
% Now, by computing $F(s)$ \eqref{Eq02} for a variety of time scales $s$, an understanding of long-range correlations may obtained since the exponent of long-range correlations $\alpha$ decline as a power law
\begin{equation}\label{Eq03}
	\small{F_m^{'}(s) = \sqrt{\frac{1}{2N_s}\sum_{v=1}^{2N_s} \Big(\frac{1}{s}\sum_{i=1}^{s}\sum_{n=1}^{m}\left(y_{i_n}-\tilde{y}_{i_n}\right)^2\Big)},}
\end{equation}
where $y_{i_n}$ and $\tilde{y}_{i_n}$ respectively denote the profile and polynomial fit for the $n$th channel. Observe from \eqref{Eq03} that the Euclidean norm of each $m$-variate error observation is used to formulate a multichannel fluctuation function $F_m^{'}(s)$ in \cite{xiong20171stMDFA} which completely disregards the cross-channel 
correlations in the data and leads to spurious long range correlation scores.
\vspace{-3mm}
\section{Proposed Methodology}
\vspace{-2mm}
\label{sec:guidelines}
This section outlines our proposed multiscale multivariate signal denoising method. For this purpose, we first describe the proposed generic multichannel extension of DFA that underpins our denoising framework.
%develop a novel MDFA that underpins our denoising framework by incorporating the cross-channel correlations of multivariate data and is explained first.
%The advent of MVMD presents a case of extension of this technique to multivariate signal denoising but this extension is not trivial

%Hence, a time series composed mostly on random noise can be identified by based on its long-range correlation analysis using DFA.
\vspace{-3mm}
%This section first describes a novel multichannel DFA (MDFA) designed specifically to underpin a multivariate denoising framework based on MVMD which is explained later.
\begin{figure}
	\centerline{\includegraphics[width=0.68\columnwidth]{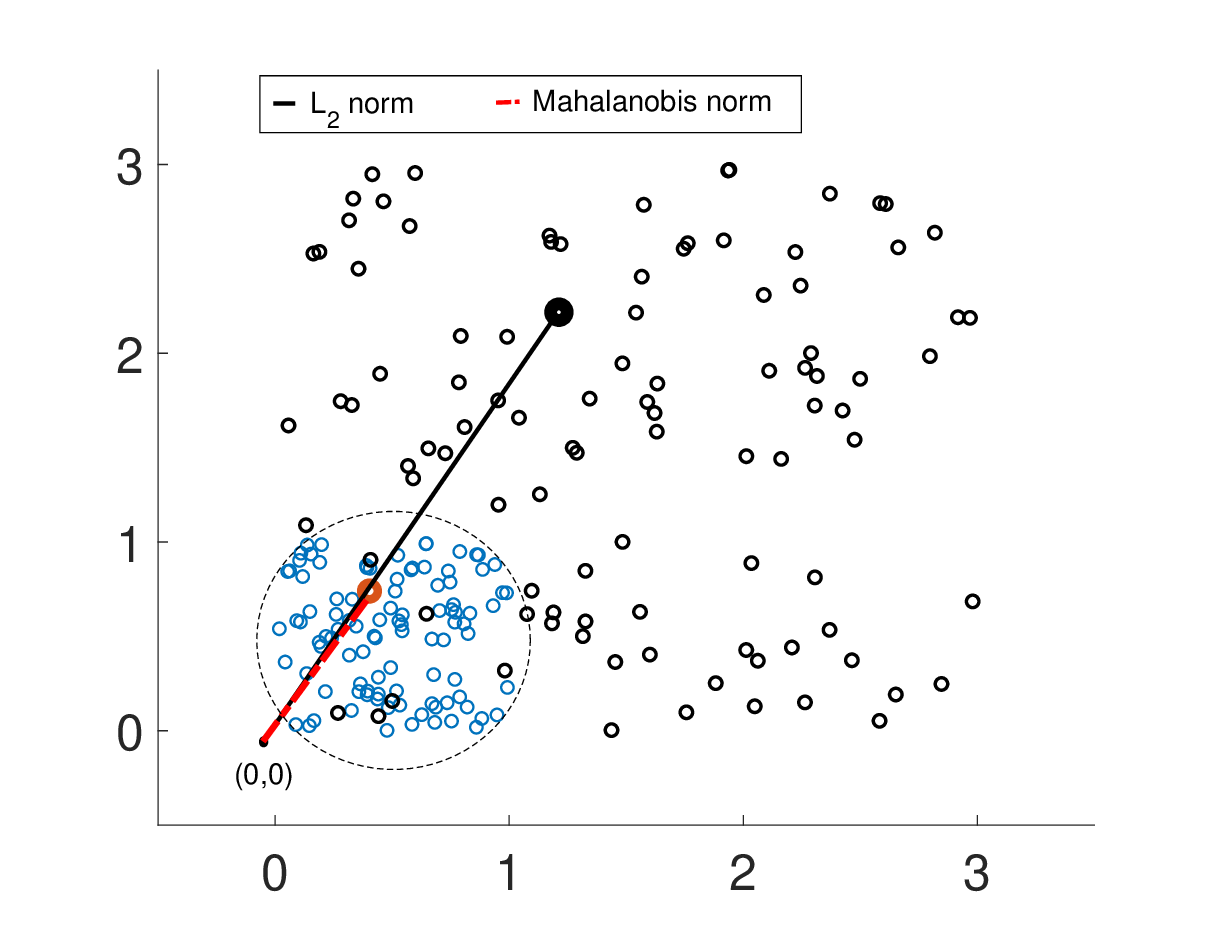}}
	\vspace{-4mm}
	\caption{\small{Relocation of a set of points in 2D space (black circles) to another set (blue circles) through Mahalanobis distance norm. The black line shows the standard euclidean norm of a single point in 2D; the red line shows the corresponding Mahalanobis norm of the same point, which is more accurate since it considers cross-correlations and is robust to variance bias across the channels of multivariate data.} 
		%		On the contrary, ED (depicted using the dark black line) fails to consider the cross-correlations and is sensitive to variance bias, i.e., ED changes with change in variance of a dataset but MD remains unchanged.
	}
	\label{fig01}
	\vspace{-5mm}
\end{figure} 
%\vspace{-1mm}
\subsection{A Generic Multichannel Extension of DFA}
\vspace{-1mm}
%The existing multichannel DFA \cite{xiong20171stMDFA} completely disregards the cross-channel correlations of a multivariate time series and can lead to spurious scores especially for cross-correlated multichannel data. 
To address the aforementioned weakness in the existing DFA in [14], we propose a novel multichannel extension of the DFA method that considers cross-correlations via Mahalanobis distance (MD) measure and may be seen as a generalization of \cite{xiong20171stMDFA}. The steps involved in the proposed Generic Multichannel DFA, termed GMDFA, are given below:

Given a multivariate time series $\pmb{x}_i \ \forall \ i=1,...,N$, where $\pmb{x}_i=[x_{i_1},\ldots,x_{i_m}]^T\in\mathcal{R}^m$ represents an $m$-variate observation at time index $i$, the cumulative sum $\pmb{y}_i$ is computed via
\vspace{-1mm}
\begin{equation}\label{Eq04}
	\small{\pmb{y}_i = \frac{1}{N}\sum_{i=1}^N (\pmb{x}_i-\pmb{\overline{x}}),}
	\vspace{-2mm}
\end{equation}
where $\pmb{\overline{x}}=\frac{1}{N}\sum_{i=1}^{N} \pmb{x}_i$ denotes the multichannel mean.
%$\{\pmb{y}_i\ \forall\ i=(v-1)\times s,\ldots,v\times s\}$where $v=1,\ldots,N_s$

Next, the signal $\pmb{y}_i =y_{i_n}\}_{n=1}^m$ for all $i =1,\ldots, N$, is divided in $2N_s$ spatial segments by cutting it into $N_s=N/s$ segments of equal lengths starting from both ends of the series. Then, the local trend $\tilde{\pmb{y}}_i=\tilde{y}_{i_n}\}_{n=1}^m$ is estimated based on the quadratic polynomial fit of each channel
\begin{equation}\label{Eq05}
	\tilde{y}_{i_n} = a_n \cdot i^2 + b_n \cdot i + c_n, \ \ \ \ i=1,\cdots,s,
\end{equation}
where $a_n, b_n, c_n$ denote the coefficients required for the least square fit $\tilde{y}_{i_n}\sim y_{i_n}$. Here, quadratic polynomial is used to estimate the slowly varying background trend. We next provide the mathematical definition of Mahalanobis norm which forms the basis of our proposed method. 
%albeit a higher order polynomial fit could also be used.

% which can be rewritten as
%\begin{equation}
%\|\Sigma^{-\frac{1}{2}}\left(\pmb{y}_{i}-\tilde{\pmb{y}}_{i}\right)\|_2^2 =\sum_{n=1}^m w_i (y_{i_n}-\tilde{y}_{i_n})^2
%\end{equation}
% where the weights $w_i$ are based on the covariance matrix $\Sigma^{-1}$.
\begin{definition}[Mahalanobis norm]
	Let $\Sigma$ denote a symmetric and positive definite covariance matrix of vector observations $\pmb{z}_{i}\}_{i=1}^N$, we define the Mahalanobis norm $\|\pmb{z}_i\|_\Sigma= \sqrt{\pmb{z}_{i}^T\Sigma^{-1} \pmb{z}_{i}}$ that satisfies the following properties of a norm on that vector space $\mathcal{Z}$, i.e.,
	\begin{enumerate}
		\item $	\|\pmb{z}\|_\Sigma > 0 \ \forall \ \pmb{z}\ne\pmb{0}$;
		\vspace{-2mm}
		\item $	\|\pmb{z}\|_\Sigma =0$ \textit{iff} $\pmb{z}=\pmb{0}$;
		\vspace{-2mm}
		\item $	\|a\pmb{z}\|_\Sigma = |a|\cdot \|\pmb{z}\|_\Sigma$ for a scalar $a$;
		\vspace{-2mm}
		\item $	\|\pmb{z}_1+\pmb{z}_2\|_\Sigma \leq \|\pmb{z}_1\|_\Sigma + \|\pmb{z}_2\|_\Sigma$.
	\end{enumerate}
	where the vectors $\pmb{z}$, $\pmb{z}_1$ and $\pmb{z}_2$ belong to the space $\mathcal{Z}$. 
%	The proof of these properties are provided as a supplement material with this manuscript.
\end{definition}
\begin{remark} Mahalanobis norm $\|\pmb{z}_i\|_\Sigma$ is a generalized multivariate norm because (a) it considers cross channel dependencies which are completely ignored within the $L_2$ norm; and (b) it performs variance normalization to remove variance bias across the channels (as depicted in Fig \ref{fig01}). 
	\label{remark1}
\end{remark}	
That can be observed from the following two cases of uncorrelated multivariate data where $\|\pmb{z}_i\|_\Sigma$ reduces to a form of $\|\pmb{z}_i\|_2$. Firstly, when $\tiny{\Sigma=I_{m\times m}}$ that denotes an identity matrix, $\|\pmb{z}_{i}\|_\Sigma$ is given by
	\vspace{-1mm}
	\begin{equation}\label{Eq06}
		\small{\|\pmb{z}_{i}\|_{\tiny{\Sigma=I_{m\times m}}}= \sqrt{\pmb{z}_{i}^T I_{m\times m}^{-1}\pmb{z}_{i}}=\sqrt{\pmb{z}_{i}^T \pmb{z}_{i}}=\|\pmb{z}_{i}\|_2}.
		\vspace{-1mm}
	\end{equation}
	%\label{remark1}
	%where $\pmb{z}_i=[z_{i_1}, \ldots,z_{i_m}]^T\in \mathcal{R}^m$ is a vector observation.
	%\end{remark}
	%An implication of \textit{Remark} \ref{remark1} is that if multivariate data is uncorrelated and all its channels are unit variance then $L_2$ norm and Mahalanobis norm are equal.
	%\begin{remark} 
	
\noindent Secondly, when $\Sigma=\boldsymbol{\sigma}^T I_{m\times m}$ is a diagonal matrix where the vector $\boldsymbol{\sigma} = [\sigma_1,\sigma_2,\ldots,\sigma_m]^T$ contains channel variances, $\|\pmb{z}_{i}\|_\Sigma$ is given by
	%the Mahalanobis norm is nothing but $L_2$ norm of variance normalized observations $\overline{\pmb{z}}_i=[\frac{z_{i_1}}{\sigma_1},\ldots,\frac{z_{i_m}}{\sigma_m}]^T$ as can be observed from the following bivariate expansion 
	%	  with $\tiny{\Sigma= 
	%			 	\begin{pmatrix}
	%			 	\sigma_1^2 & 0 \\
	%			 	0 & \sigma_2^2
	%			 	\end{pmatrix}}$, 
	%		 scalar expansion of $\|\pmb{z}_{i}\|_{\Sigma}$ is given as follows
	%
	%{\tiny{\Sigma=\begin{pmatrix}
	%			\sigma_1^2 & 0 \\
	%			0 & \sigma_2^2
	%\end{pmatrix}}}
	\begin{equation}\label{Eq07}
		\small{
			\|\pmb{z}_{i}\|_{\tiny{\Sigma=\boldsymbol{\sigma}^T I_{m\times m}}}
			=\sqrt{\left(\frac{z_{i_1}}{\sigma_1}\right)^2 + \ldots + \left(\frac{z_{i_m}}{\sigma_m}\right)^2} = \|\overline{\pmb{z}}_i\|_2},
		\vspace{-1mm}
	\end{equation}
	where $\pmb{z}_i=[z_{i_1}, \ldots,z_{i_m}]^T$ and  $\overline{\pmb{z}}_i=[\frac{z_{i_1}}{\sigma_1},\ldots,\frac{z_{i_m}}{\sigma_m}]^T$.
	%	\label{remark2}
	%
	%
	%\begin{remark}
	\begin{figure}[t]
		\centering
%		\begin{subfigure}{.3\textwidth}
%			%		\centering
%			\includegraphics[width=\linewidth]{PropMDFAvPrevMDFA.eps}
%			%		\caption{Noisy at 10 dB}
%			%		\label{HvyDopp}
%		\end{subfigure}
		%		\hspace{-6mm}
			\begin{subfigure}{.27\textwidth}
				%		\centering
				\includegraphics[width=\linewidth]{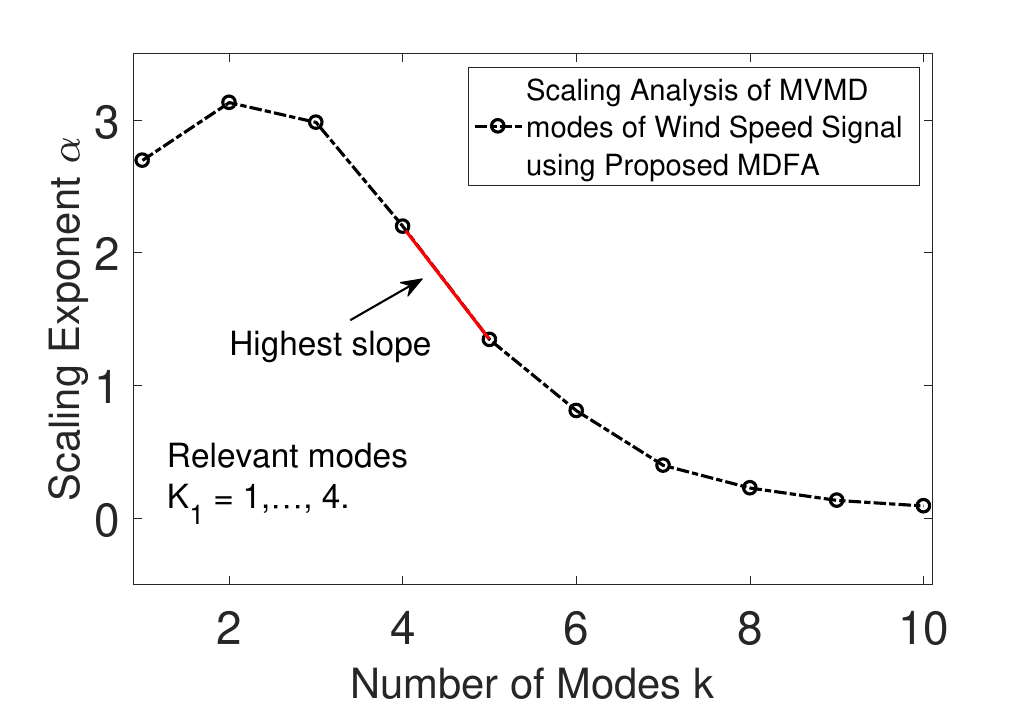}
				%		\caption{Noisy at 10 dB}
				%		\label{HvyDopp}
			\end{subfigure}
		\caption{\small{Plot of scaling exponents $\alpha_k$, computed using proposed GMDFA, for MVMD modes of noisy Wind signal at $10$ dB.}}
		\label{Fig02}
		\vspace{-5mm}
	\end{figure}

Finally, in the case of correlated multivariate data, Mahalanobis norm essentially computes the $L_2$ norm by un-correlating the variance normalized vector observations as depicted in Fig. \ref{fig01}. For a special case of bivariate data, $\|\pmb{z}_{i}\|_\Sigma$ can be rewritten as
	\begin{equation}\label{Eq08}
		\small{
			\|\pmb{z}_{i}\|_\Sigma
			=\frac{1}{\sqrt{1-\rho^2}}\sqrt{\|\overline{\pmb{z}}_i\|_2^2 - \frac{2\rho z_{i_1} z_{i_2}}{\sigma_1 \sigma_2}}},
		\vspace{-2mm}
	\end{equation}
	\label{remark3}
	%\end{remark}
where $\rho$ denotes correlation coefficient and $\Sigma = \tiny{ 
		\begin{pmatrix}
			\sigma_1^2 & \rho \sigma_1 \sigma_2 \\
			\rho \sigma_1 \sigma_2 & \sigma_2^2
	\end{pmatrix}}$.
	%first uncorrelated the multivariate observations and then computes the $L_2$ norm of variance normalized observations $\overline{\pmb{z}}_i$.
	
%	\vspace{-3mm}

Based on \textit{remark} \ref{remark1}, we utilize the generic (Mahalanobis) norm $\|\pmb{y}_{i}-\tilde{\pmb{y}}_{i}\|_\Sigma$ to formulate a purely multivariate fluctuation function $F_m^\Sigma(s)$ within MDFA, that is,
\begin{equation}\label{Eq09}
	\small{F_m^\Sigma(s) = \sqrt{\frac{1}{2sN_s}\sum_{v=1}^{2N_s} \sum_{i=vs+1}^{(v+1)s}\left(\pmb{y}_{i}-\tilde{\pmb{y}}_{i}\right)^T\Sigma^{-1} \left(\pmb{y}_{i}-\tilde{\pmb{y}}_{i}\right)}},
\end{equation}
where covariance matrix $\Sigma$ characterizes the interchannel dependencies within the detrend (or fluctuations) $\pmb{y}_{i}-\tilde{\pmb{y}}_{i}$. 

It is clear that \eqref{Eq03} becomes a special case of \eqref{Eq09} for identity covariance matrix, i.e., uncorrelated input multichannel data. For more interesting cases involving multichannel data that exhibit cross-channel correlations, \eqref{Eq09} provides more informative fluctuation scores.
%\begin{figure}[t]
%	\centering
%	\begin{subfigure}{.24\textwidth}
%		%		\centering
%		\includegraphics[width=\linewidth]{PropMDFAvMDFA.eps}
%		%		\caption{Noisy at 10 dB}
%		%		\label{HvyDopp}
%	\end{subfigure}
%	\hspace{-5mm}
%	\begin{subfigure}{.24\textwidth}
%		%		\centering
%		\includegraphics[width=\linewidth]{ZoomedPropMDFAvMDFA.eps}
%		%		\caption{Noisy at 10 dB}
%		%		\label{HvyDopp}
%	\end{subfigure}
%	
%	%	\hspace{-5mm}
%	\begin{subfigure}{.24\textwidth}
%		%		\centering
%		\includegraphics[width=\linewidth]{Wind_PreviousMDFA_10dB.eps}
%		%		\caption{MMD (Previous MDFA)}
%		%		\label{WLS}
%	\end{subfigure}
%	\hspace{-5mm}
%	\begin{subfigure}{.24\textwidth}
%		\centering
%		\includegraphics[width=\linewidth]{Wind_ProposedMDFA_10dB.eps}
%		%		\caption{MDD (Prop. MDFA)}
%		%		\label{HvyDopp}
%	\end{subfigure}
%	%	\hspace{-5mm}
%	%	\begin{subfigure}{.25\textwidth}
%	%		\centering
%	%		\includegraphics[width=\linewidth]{Sofar_MMD_10dB.eps}
%	%		%		\caption{Prop. MDD}
%	%		\label{WLS}
%	%	\end{subfigure}
%	%	\vspace{-6}
%	\caption{\small{Comparison of denoising performance of the proposed MDFA against the existing multichannel DFA \cite{xiong20171stMDFA}.}
%		%		 when used within the proposed denoising framework based on MVMD.
%	}
%	\label{Fig03}
%	\vspace{-6mm}
%\end{figure}

In order to perform multichannel scaling analysis in \eqref{Eq09}, $F_m^{\Sigma}(s)$ is computed for varying time scales where generally the range $s=4,\ldots,16$ is used \cite{kantelhardt2001detecting}. Finally, a scaling exponent $\alpha$ is computed using power law representation of $F_m^\Sigma(s)$
\begin{equation}\label{Eq10}
	\vspace{-2mm}
	F_m^\Sigma(s) = s^{\alpha}.
%	\vspace{-1mm}
\end{equation}
In practice, $\alpha$ is calculated based on the slope of the plot between $ln F_m^\Sigma(s)$ and $ln\ s$ because $log_s F_m^\Sigma(s)=\frac{ln F_m^\Sigma(s)}{ln\ s}$, where $ln$ denotes the natural logarithm operator.
\vspace{-3mm}
\subsection{Multiavriate Denoising Using MVMD and GMDFA}
Here, we present a multivariate signal denoising method that applies the proposed GMDFA on the data-driven modes of noisy signal obtained from MVMD, as discussed below:
\vspace{-3mm}
\subsubsection{Multiscale decomposition using MVMD} 
%\vspace{-4mm}
%\section{Multivariate Variational Mode Decomposition}
\vspace{-1mm}
%Recently, a multivariate extension of the variational mode decomposition (VMD) is proposed in 
Multivariate VMD \cite{ur2019MVMD} is a generic multichannel extension of the VMD algorithm that decomposes a multivariate signal $\pmb{x}_i \in\mathcal{R}^m$ into $K$ number of predefined multivariate modulated oscillations $\pmb{u}_{k,i}\in\mathcal{R}^m$ which are based on a common frequency component across all channels.
\vspace{-1mm} 
\begin{equation} \label{Eq11}
	\small{\pmb{x}_{i_n}=\sum_{k=1}^K \pmb{u}_{k,i_n}}.
\end{equation}
%The objective function within MVMD has been defined as the bandwidth of $K$ number of multivariate modulated oscillations across all channels. That is subject to the multiple constraints of signals across multiple channels being equal to the sum of the decomposed components. The resulting optimization problem is given below.  
%%The cost function of multivariate VMD (MVMD) is obtained by taking the Frobeneus norm, as a direct multichannel extension of the $L_2$ norm used in VMD, as follows
%\vspace{-2mm}
%\begin{equation}\label{Eq03}
%	\small{\mathop{\text{minimize}}_{u_{k,n},w_k}\left \{ \sum_{k=1}^K\sum_{n=1}^m\left \| \partial_t \left[u_{k,i_n}^{+}\ e^{-jw_k i}\right] \right \|_2^2\right \},}
%\end{equation}
%where $u_{k,i}^n$ denote multichannel band-limited intrinsic mode functions (BLIMFs) represented as a function of modulated multivariate oscillation $u_{k,i_n}^{+}$, assuming a common frequency component $w_k=\frac{d\phi_k}{dt}$.
%%as follows
%%\begin{equation}\label{Eq02}
%%u_{k,i}^{n+} = u_{k,i}^n+j\mathcal{H}u_{k,i}^n=|u_{k,i}^{n+}|\ e^{j\phi_k t} \ \forall \ n=1,\ldots,m,
%%\end{equation}
%%where $\mathcal{H}u_{k,i}^n$ denotes Hilbert transform of $u_{k,i}^n$.
%Owing to that MVMD avoids the mode mixing problem otherwise observed in MEMD that makes MVMD useful in a variety of applications including denoising. 
\begin{figure}[t]
	\centering
	\begin{subfigure}{.25\textwidth}
		%		\centering
		\includegraphics[width=\linewidth]{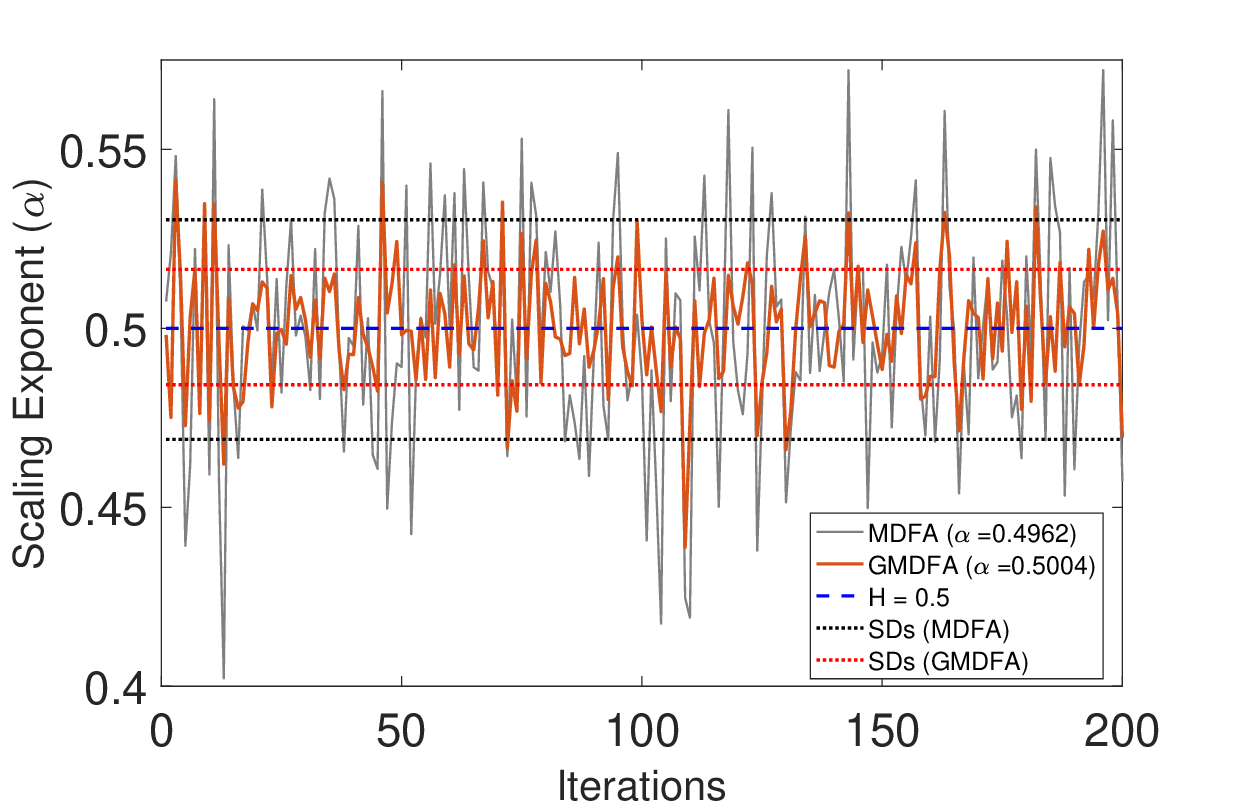}
		%		\caption{Noisy at 10 dB}
		%		\label{HvyDopp}
	\end{subfigure}
	\hspace{-5mm}
	\begin{subfigure}{.25\textwidth}
		%		\centering
		\includegraphics[width=\linewidth]{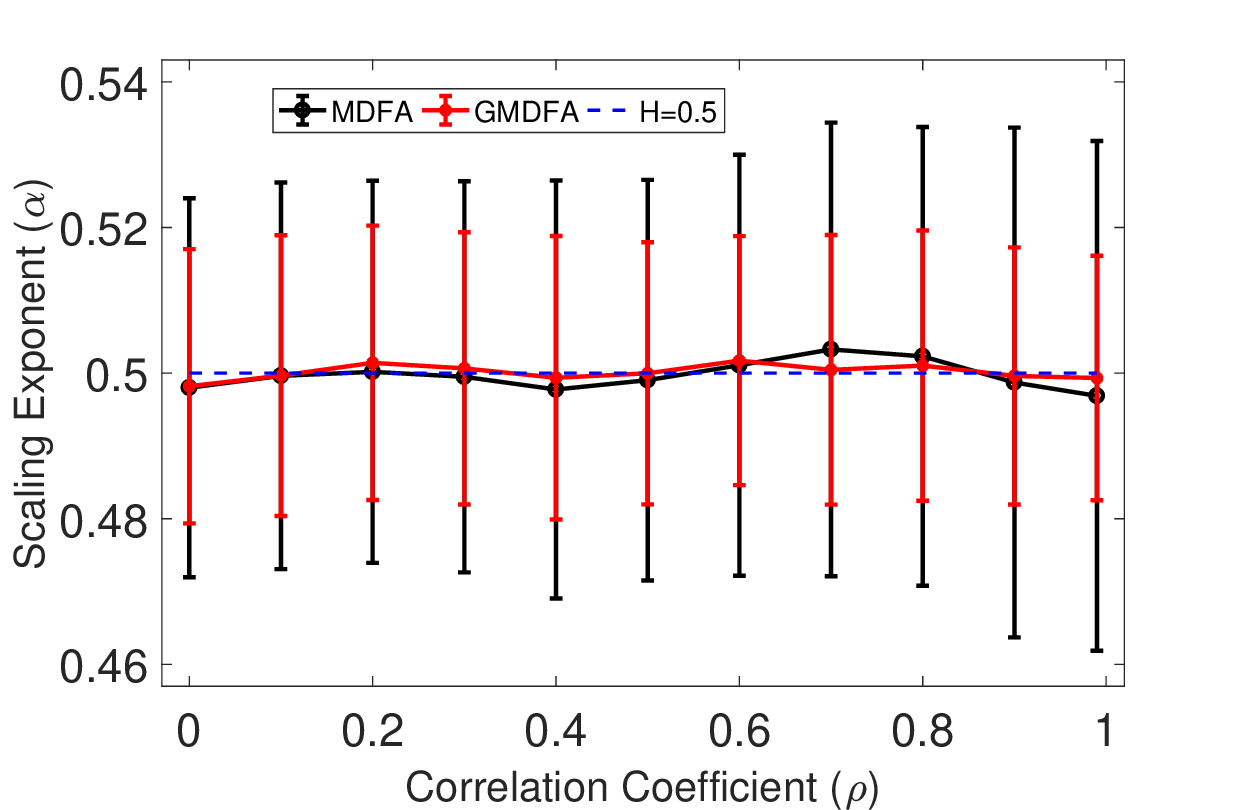}
		%		\caption{Noisy at 10 dB}
		%		\label{HvyDopp}
	\end{subfigure}
	
	%	\begin{subfigure}{.32\textwidth}
	%		%		\centering
	%		\includegraphics[width=\linewidth]{TrivariateRho23VAlpha_wGnh5.eps}
	%		%		\caption{Noisy at 10 dB}
	%		%		\label{HvyDopp}
	%	\end{subfigure}
	\caption{\small{Numerical simulations on bivariate white gaussian noise (wGn) for estimation of the Hurst exponent $H=0.5$ using MDFA \cite{xiong20171stMDFA} and GMDFA at varying degrees of cross-correlation.}}
	\label{wGnAlpha}
	\vspace{-5mm}
\end{figure}

Within our proposed denoising approach, firstly MVMD is used to decompose a noisy multivariate signal $\pmb{x}_i$ into an ensemble of $K$ multichannel BLIMFs $\pmb{u}_{k,i}$ which comprise of modulated multivariate oscillations of a common frequency component. 
Among those, initial BLIMFs contain low frequency (or smooth) oscillations whereas the latter BLIMFs mostly comprise of high frequency fluctuations. This representation can be mathematically written as
\vspace{-1mm}
\begin{equation}\label{Eq12}
	\small{\pmb{x}_{i}=\sum_{k=1}^{K} \pmb{u}_{k,i} =\sum_{k=1}^{K_1} \pmb{u}_{k,i} + \sum_{k=K_1+1}^K \pmb{u}_{k,i}}, \ \forall\ i=1,\ldots,N,
	\vspace{-2mm}
\end{equation}
where $\{\pmb{u}_{k,i}\}_{k=1}^{K_1}$ denotes the set of initial BLIMFs containing majority of (true) signal and $\{\pmb{u}_{k,i}\}_{k=K_1+1}^{K}$ denotes the BLIMFs with predominant noise. Next, MDFA is used to detect the predominant noise modes, i.e., $K_1$.  

%is obtained we our method aims to partially reconstruct the denoised signal based on these initial relevant signal modes and reject the subsequent modes with predominant noise. In results in smooth (or low frequency) oscillations of signal into a few initial multichannel BLIMFs while high frequency (signal) fluctuations and noise are distributed across latter modes, as follows
\vspace{-3mm}
\subsubsection{Rejection of predominantly noisy BLIMFs using MDFA}
The proposed GMDFA is used to identify and discard predominantly noisy BLIMFs based on (a) their higher frequency content and (b) absence of long-range auto-correlations. 
% in the multivariate observations $\pmb{u}_{k,i}$, within the $k$th BLIMF, are independent (i.e., no auto-correlations) with highly varying (high-frequency) structure.
%Therefore, detrended fluctuation analysis may be used to identify since it informs about the frequency of the individual mode. This insight into the statistics of multichannel modes is to effectively differentiate between the initial signal modes $\{\pmb{u}_{k,i}\}_{k=1}^{K_1}$ and latter noise modes $\{\pmb{u}_{k,i}\}_{k=K_1+1}^{K}$.
%In this regard, use of proposed MDFA is motivated by the fact that multich cross-correlated and to get true auto-correlation score these in that case. 
%in order to obtain the representation \eqref{Eq11} from the ensemble of modes obtained by solving \eqref{Eq01}. 
%For this purpose, we propose to use our MDFA because it incorporates the cross-channel-correlations observed in noisy multivariate data.
In this regard, the comparative analysis of the scaling exponents $\alpha_k$, computed for each BLIMF $\pmb{u}_{k,i}$ using \eqref{Eq12}, is performed. 
%For instance,  where initial BLIMFs with smooth signal oscillations yield higher $\alpha_k$ while subsequent BLIMFs with higher frequency components yield lower value for $\alpha_k$.  
Understandably, $\alpha_k$ should decrease for every higher order BLIMF of the MVMD owing to the presence of increasingly high frequency fluctuations and decreasing long-range correlations; that is evident from Fig. \ref{Fig02} that plots $\alpha_k$ for MVMD modes of a noisy trivariate wind signal.
%This type of representation allows to differentiate between signal and noise modes by observing the rate of change of scaling exponent when moving from one mode to the other.

Let $\beta_k$ denote the slope of the line connecting the exponents $\alpha_k$ and $\alpha_{k+1}$ for two consecutive modes, i.e.,
\vspace{-1mm}
\begin{equation}\label{Eq13}
	\small{\beta_k = |\alpha_{k+1} - \alpha_k|}.
	\vspace{-2mm}
\end{equation}

Then, $\beta_k$ quantifies the amount of change in the frequency of the fluctuations (or decrease in long-range correlations) when moving one mode to the other. That means, highest slope suggests maximum increase in frequency or maximum decrease in long-range correlations, i.e., largest increase in noise content. Consequently, the first mode after the highest slope, i.e., $\pmb{u}_{K_1+1}$, marks the beginning of predominantly noisy modes where $K_1$ may be computed as follows
\vspace{-1mm}
\begin{equation}\label{Eq14}
	\small{K_1 = \mathop{\text{argmax}}_{k}\{\beta_1,\ldots,\beta_K\}}.
	\vspace{-1mm}
\end{equation}
Subsequently, the modes $\{\pmb{u}_{k,i}\}_{k=K_1+1}^{K}$ are rejected as noise.
%\vspace{-1mm}  
\subsubsection{Reconstruction} The remaining multichannel BLIMFs $\{\pmb{u}_{k,i}\}_{k=1}^{K_1}$, corresponding to relevant signal, may contain traces of noise which are removed by applying principal component analysis (PCA) separately on each multichannel mode, as suggested in \cite{aminghafari2006MWD}. 
%In this regard, we employed heuristic approach \cite{karlis2003PCA_SelectionRule} to identify the principal components. Finally, 
Following the application of PCA \cite{karlis2003PCA_SelectionRule}, the denoised multivariate signal is obtained based on the post-processed selected relevant modes $\{\breve{\pmb{u}}_{k,i}\}_{k=1}^{K_1}$, as follows
\vspace{-1mm}
\begin{equation}\label{Eq15}
	\small{\hat{\pmb{s}}_{i}=\sum_{k=1}^{K_1} \breve{\pmb{u}}_{k,i} \ \ \ \forall\ \  i=1,\ldots,N,}
	\vspace{-1mm}
\end{equation}
where $\hat{\pmb{s}}_{i}$ denotes the denoised multivariate signal.
%Observe that the use of covariance matrix (to characterize the cross-correlations) in our MDFA leads to significantly different scaling exponent $\alpha$ scores compared to that of \cite{} for correlated data. That may have significant impact on the decisions based on these scores specifically when detecting BLIMFs with cross-correlated noise/ signal.
\begin{table}[t]
	\caption{Input versus output SNR values of various comparative multivariate signal denoising methods on real signals.}
	\vspace{-2mm}
	\centering
	\scalebox{0.485}{
		\resizebox{\textwidth}{!}{
			\setlength\extrarowheight{1pt}
			\begin{tabular}{|c||c|c|c|c||c|c|c|c||c|c|c|c|}\hline
				\textbf{Avg. In. SNR}
				%				&\multicolumn{4}{c}{\textbf{-10}} 
				&{\textbf{-2}} &{\textbf{2}} &{\textbf{6}} &{\textbf{10}} &{\textbf{-2}} &{\textbf{2}} &{\textbf{6}} &{\textbf{10}} &{\textbf{-2}} &{\textbf{2}} &{\textbf{6}} &{\textbf{10}}\\
				\hline
				\textbf{Test Signal}
				&\multicolumn{3}{c}{\textbf{Bi. Sofar Signal}} & &\multicolumn{3}{c}{\textbf{Tri. Wind Signal}} & &\multicolumn{3}{c}{\textbf{Qd. Synthetic Signal}}&\\
				\hline 
				\textbf{MWD} \ \ \ \ \ bal.
				%				&3.84  &&  3.98  & & 3.91 
				& 6.86 & 11.11 & 14.56 & 18.66 & \textbf{9.13} & 11.25 & 12.07 & 12.99 & 6.69 & 10.33 & 13.62 & 17.05\\
				%				&\textbf{3.24} && 3.95 && 3.60 
				\ \ \ \ \ \ \ \ \ \ \ \ unbal. & 6.50 & 10.93 & 14.58 & 18.27 & \textbf{8.86} & 10.71 & 11.89 & 12.77 & 6.55 & 10.23 & 13.80 & 16.75\\
				\hline
				\textbf{MWSD} \ \ \ \ bal.
				%				&-4.13 && -5.65 &&-4.89 
				& 1.86 & 2.93 & 3.65 & 4.28 & 0.28 & 0.75 & 0.94 & 1.01 & 3.76 & 5.06 & 5.64 & 5.90\\
				%				&-4.71  && -4.76 && -4.74 
				\ \ \ \ \ \ \ \ \ \ \ \ unbal.& 1.51 & 2.52 & 3.42 & 4.05 & 0.18 & 0.70 & 0.89 & 0.99 & 3.06 & 4.35 & 5.42 & 5.73\\
				%				\hline
				%				\textbf{MEMD} \ \ \ bal.
				%				%				& 1.10 && 0.77 && 0.93 
				%				& 5.44 & 5.97 & 6.90 & 8.78 & 2.62 & 3.17 & 6.77 & 9.00 \\
				%				%				& 0.36 && 3.88 && 2.12 
				%				\ \ \ \ \ \textbf{-IT} \ \ \ \ \ unbal.& 7.24 & 7.08 & 7.67 & 8.65 & -0.04 & 4.30 & 4.99 & 7.82\\
				\hline
				\textbf{MMD} \ \ \ \ \ bal.
				%				&\textbf{5.05} & & \textbf{5.50} &  & \textbf{5.27} 
				& 7.54 & 12.20 & 15.46 & 18.94 &7.33 & 10.57 & 13.35 & 16.50 & 7.22 & 10.58 & 13.89 & \textbf{17.12}\\
				%				&1.78 && \textbf{5.74} && \textbf{3.76} 
				\ \ \ \ \ \ \ \ \ \ \ \ unbal.& 8.05 & 11.72 & 15.03 & 18.91  & 7.54 & 10.61 & 13.56 & 16.26 & 7.78 & 10.47 & 13.75 & \textbf{16.92}\\
				\hline
				\textbf{MDD} \ \ \ \ \ bal.
				%				&\textbf{5.05} & & \textbf{5.50} &  & \textbf{5.27} 
				& \textbf{8.39} & \textbf{12.65} & \textbf{16.27} & \textbf{20.22} & 8.49 & \textbf{11.69} & \textbf{15.26} & \textbf{16.95} & \textbf{8.20} & \textbf{11.83} & \textbf{14.24} & 16.76\\
				%				&1.78 && \textbf{5.74} && \textbf{3.76} 
				\ \ \ \ \ \ \ \ \ \ \ \ unbal.& \textbf{8.56} & \textbf{12.02} & \textbf{16.50} & \textbf{19.38} & 8.33 & \textbf{11.64} & \textbf{14.61} & \textbf{16.84} & \textbf{8.31} & \textbf{11.42} & \textbf{14.09} & 15.80\\
				\hline
	\end{tabular}}}
	\label{Table1}
	\vspace{-4mm}
\end{table} 
%\begin{figure}[t]
%	\begin{minipage}[b]{1\linewidth} \centerline{\includegraphics[scale=0.15]{Sofar_NoisyResults4MDDPaper10dB.eps}} \centerline{(a) \small{Noisy at $10$ dB}} \end{minipage}
%	\vspace{-8mm}
%	\begin{minipage}[b]{1\linewidth} \centerline{\includegraphics[scale=0.1]{Sofar_MMDResults4MDDPaper10dB.eps}} \centerline{(b) \small{MDD}} \end{minipage}\\
%	\begin{minipage}[b]{1\linewidth} \centerline{\includegraphics[scale=0.22]{Sofar_MWDResults4MDDPaper10dB.eps}} \centerline{(c) \small{MWD}} \end{minipage}
%	\begin{minipage}[b]{1\linewidth} \centerline{\includegraphics[scale=0.22]{Sofar_MMD_10dB.eps}} \centerline{(d) \small{MDD}} \end{minipage}
%	\caption{Noisy `Tai Chi' signal (a) and Denoised `Tai Chi' signals for various methods ((b) BLFDR, (c) EMD-IT, (d) DWT-GoF, (e) DT-GOF-NeighFilt, and (f) proposed VMD-CVM) for $SNR = 10 dB$.}
%	\label{Ch1:DenoisedSet2}
%\end{figure}

\vspace{-3mm}
\section{Results and Discussion}
\vspace{-3mm}
Before demonstrating the prowess of our denoising approach, we first verify the accuracy of the proposed GMDFA in estimating the true Hurst exponent of a cross-correlated bivariate data set. The input data consisted of a long (length=$2^{16}$) bivariate wGn signal for varying cross-correlation coefficient $\rho$ values. We show our results in Fig. \ref{wGnAlpha}; the sub figure on the left side shows estimated $\alpha_k$ values obtained in the first 200 iterations using both MDFA and GMDFA for a specific value of $\rho=0.5$; the subfigure on the right shows estimated scaling exponents in the form of an error bar plot, for different correlation coefficients ranging from $0 \ - \ 1$. The GMDFA provided more accurate estimates of the Hurst exponent that also exhibited lower variances across a wide range of correlation coefficients.            
%First, we demonstrate the effectiveness of the proposed GMDFA to estimate scaling exponent $\alpha$ of the cross-correlated bivariate wGn when compared to MDFA, i.e., estimate how close $\alpha$ (using GMDFA and MDFA) is to the Hust exponent $H=0.5$. In this regard, we generate bivariate wGn of size $2^{16}$ to study the impact of change in the correlation coefficient $\rho$ on the estimated exponent $\alpha$ using proposed GMDFA and MDFA, see Fig. \ref{wGnAlpha}. Here, in Fig. \ref{wGnAlpha} (left), we plot $\alpha$ values obtained in first $200$ iterations of $\alpha$ for $\rho=0.5$. While, in Fig. \ref{wGnAlpha} (left), we use error-bar plots to show how means and standard deviations (SDs) of $200$ iterations change for each $\rho=0,0.1,\ldots,0.9,1$. As can be observed proposed GMDFA yields much accurate estimate of the Hurst exponent that is also apparent from much lesser SDs in both of the given results when compared to MDFA.

Next, we evaluate the performance of the proposed multivariate denoising method using DFA, termed MDD in the sequel, against the established state of the art methods which include multivariate wavelet denoising (MWD) \cite{aminghafari2006MWD}, multivariate synchrosqueezing wavelet denoising (MWSD) \cite{ahrabian2015MWSD} and multivariate denoising based on Mahalanobis distance (MMD) \cite{ur2019MMD}. The input datasets used in our experiments include bivariate Sofar signal \cite{richardson1989estbsn}, 
%trivariate recordings of roll, pitch and yaw movements of the arm during weight lifting exercise, 
a trivariate wind speed signal and a quadrivariate synthetic signal composed of Blocks, Bumps, Doppler and Heavy-Sine signals. 
These datasets were corrupted using multivariate additive wGn and were subsequently denoised using the comparative methods.
The quality of the denoised signal is measured through the signal to noise ratio (SNR) and visual interpretation. The open source code of the MATLAB based implementation of the proposed MDD method is available online \cite{naveed2020MVMD_DFA_Matlab_Code}.

\begin{figure}[t]
	\centering
	\begin{subfigure}{.24\textwidth}
		%		\centering
		\includegraphics[width=\linewidth]{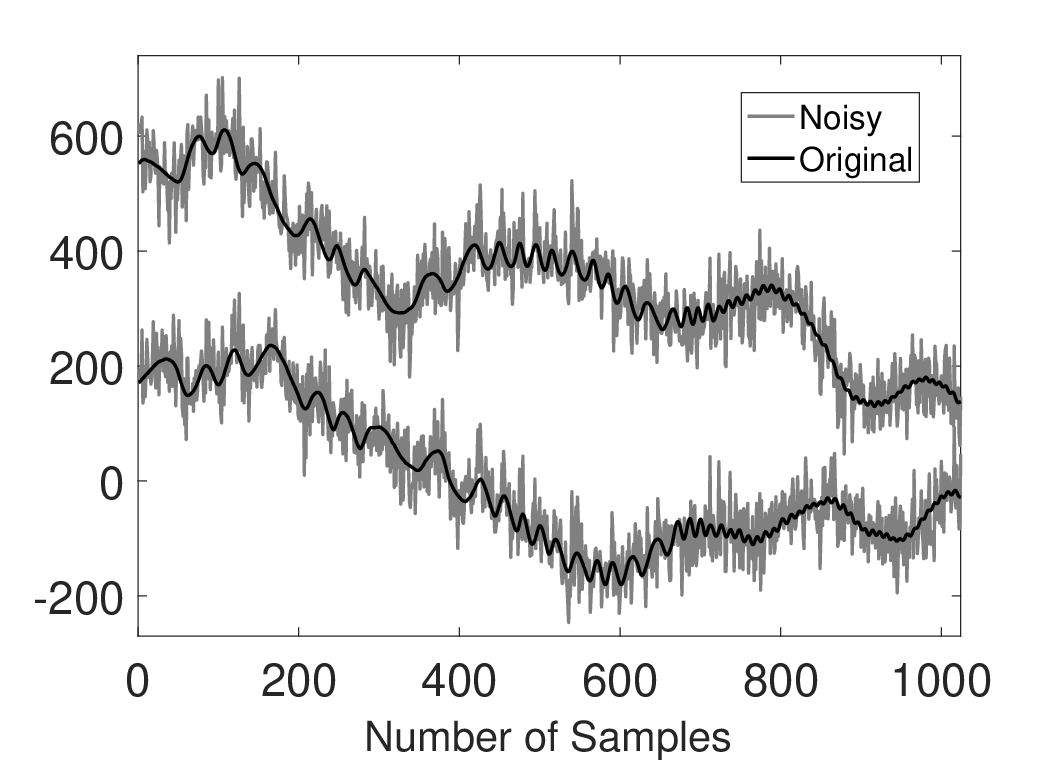}
		\caption{Noisy at 10 dB}
		%		\label{HvyDopp}
	\end{subfigure}
	\hspace{-5mm}
	\begin{subfigure}{.24\textwidth}
		%		\centering
		\includegraphics[width=\linewidth]{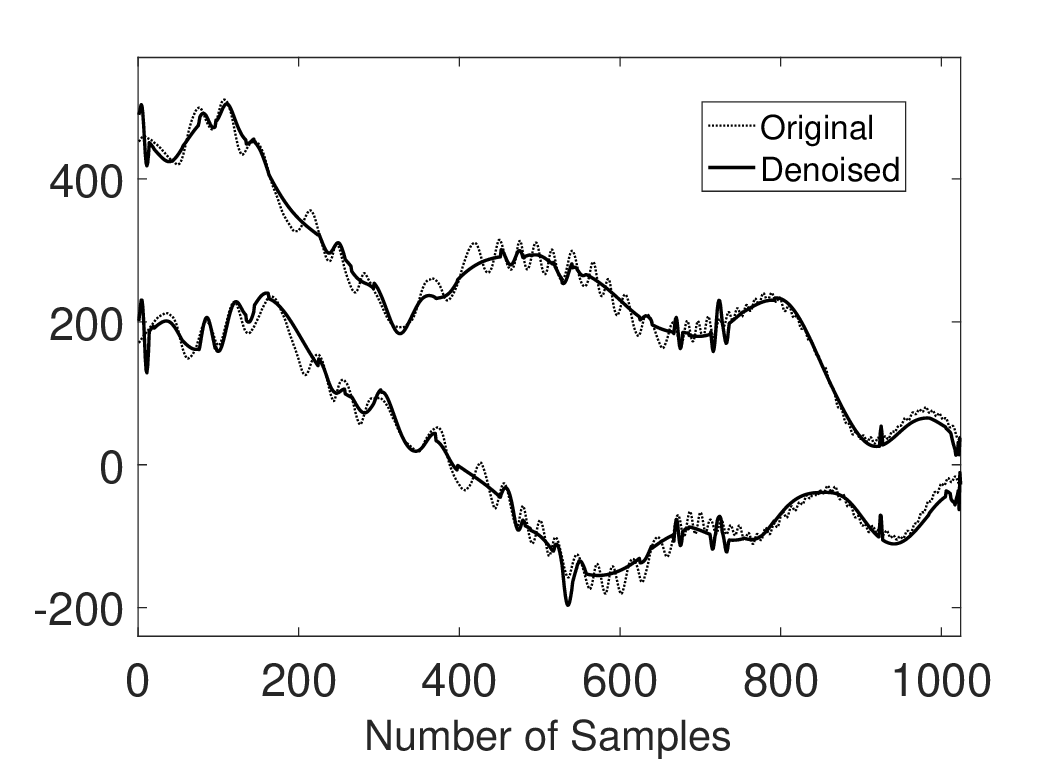}
		\caption{MMD}
		%		\label{WLS}
	\end{subfigure}
	
	%		\vspace{-1mm}
	\begin{subfigure}{.24\textwidth}
		\centering
		\includegraphics[width=\linewidth]{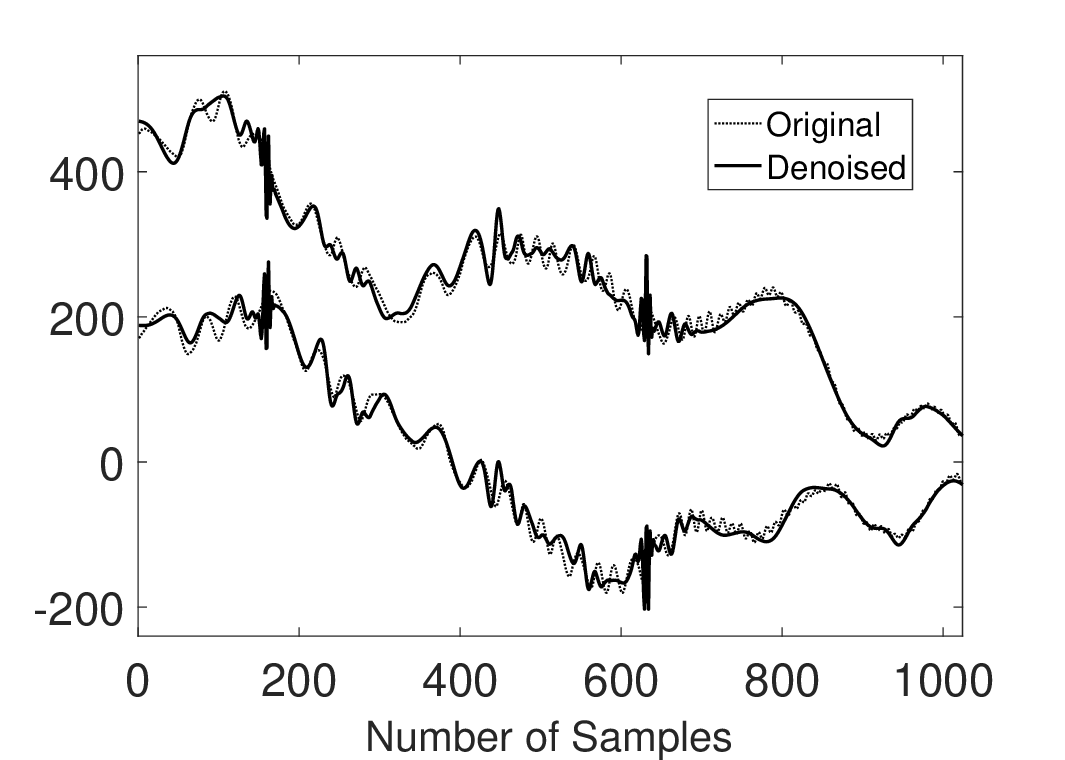}
		\caption{MWD}
		%		\label{HvyDopp}
	\end{subfigure}
	\hspace{-5mm}
	\begin{subfigure}{.24\textwidth}
		\centering
		\includegraphics[width=\linewidth]{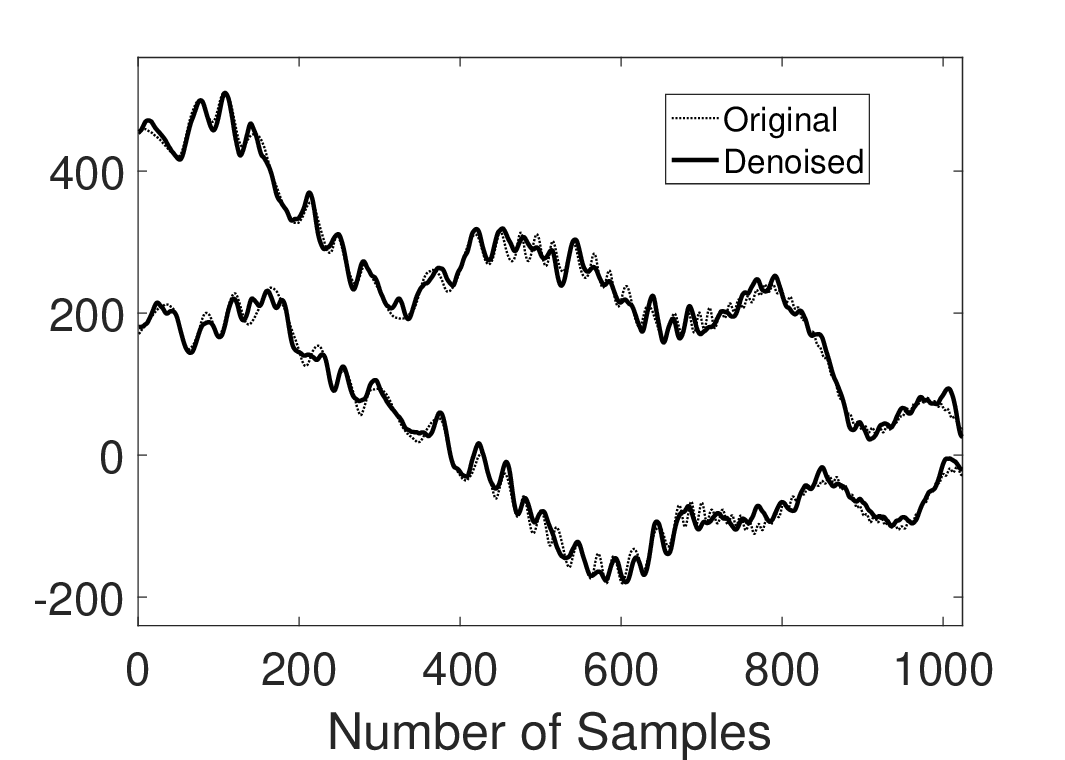}
		\caption{Prop. MDD}
		%		\label{WLS}
	\end{subfigure}
	\vspace{-2mm}
	\caption{\small{Comparison of visual denoising results of the proposed MDD method against the state of the art methods for real Sofar signal at input SNR $=10$ dB.}}
	\label{Fig04}
	\vspace{-4mm}
\end{figure}
% First, we compare the denoising performance of the proposed MDFA against the existing multichannel extension of DFA \cite{xiong20171stMDFA}. To that end, scaling exponents ($\alpha_k$) from both methods are plotted for $K=10$ BLIMFs of noisy wind signal at input SNR $10$ dB in Fig. \ref{Fig03} (top left). Observe from the zoomed-in view of the plot in Fig. \ref{Fig03} (top left) that maximum slope for \cite{xiong20171stMDFA} occurs between third and fourth mode (highlighted in dotted red line). Contrarily, using the proposed MDFA, the maximum slope is exhibited between fourth and fifth mode (highlighted in solid red line). Observe from Fig. \ref{Fig03} (bottom right) that the denoised signal reconstructed from first four BLIMFs (as suggested by the proposed MDFA) is a closer estimate of the original signal, plotted in grey in the background, compared to that plotted in Fig. \ref{Fig03} (lower left) which is reconstructed from first three BLIMFs using \cite{xiong20171stMDFA}.

Table \ref{Table1} reports average output SNRs for $J=20$ realizations from the comparative methods for all the input datasets (described above) at input SNR $=-2,2,6$ and $10$ dB. At each input noise level, we consider balanced noise (i.e., same input SNRs for all channels) and unbalanced noise cases (i.e., different input SNRs across different channels). To accentuate the best performing method, highest output SNRs are highlighted in bold for each input SNR. Observe that in most cases, the proposed MDD method yields highest output SNRs demonstrating the effectiveness of our method. Occasionally, at higher output SNRs, MMD outperforms our MDD method while MWD - generally regarded as a benchmark in multichannel signal denoising - remains competitive as well. 
%Note that a comprehensive version of Table \ref{Table1} is given in the supplement material provided with this article.

Finally, we inspect the visual quality of the reconstructed signal by displaying the denoised Sofar signals in Fig. \ref{Fig04} along with the noisy version at input SNR $=10$ dB. For meaningful qualitative analysis, we plotted original signal (shown using dotted line) in the background of the denoised signals (shown using solid line) in each case. Evidently, proposed MDD method yields best estimate of the original signal since it can estimate subtle details along with the slow variations, see Fig. \ref{Fig04} (d). On the contrary, MMD and MWD not only miss important signal details but also yield artifacts.

\vspace{-3mm}
\section{Conclusion}
\vspace{-2mm}
We have proposed a novel multivariate signal denoising method that is based on multiscale data representation and statistical signal properties. A novel and generic multichannel extension of detrended fluctuation analysis (DFA) underpins our denoising method which has been shown to outperform existing approaches owing to the full utilization of interchannel correlations within input data through utilization of Mahalanobis distance measure.

%\vfill\pagebreak

%\section{REFERENCES}
%\label{sec:refs}
%
%List and number all bibliographical references at the end of the
%paper. The references can be numbered in alphabetic order or in
%order of appearance in the document. When referring to them in
%the text, type the corresponding reference number in square
%brackets as shown at the end of this sentence \cite{C2}. An
%additional final page (the fifth page, in most cases) is
%allowed, but must contain only references to the prior
%literature.

\small
\bibliographystyle{IEEEbib}
\bibliography{ref1}

\end{document}